\documentclass[a4paper]{article}
\usepackage{amsmath}
\usepackage{amsfonts}
\usepackage[T1]{fontenc}
\usepackage[english]{babel}
\usepackage[latin1]{inputenc}
\usepackage{ae}
\usepackage{aecompl}
\newtheorem{theorem}{Theorem}
\newtheorem{lemma}[theorem]{Lemma}

\title{Static axisymmetric space-times with prescribed multipole moments}
\author{Thomas B\"ackdahl${}^*$, Magnus Herberthson\thanks{Department of  
Mathematics, Link{\"o}ping University,
SE-581 83 Link{\"oping}, Sweden.\newline
\hspace*{5mm} e-mail: thbac@mai.liu.se, maher@mai.liu.se}}
\date{}

\begin{document}
\maketitle
\begin{abstract}
In this article we develop a method of finding the static axisymmetric 
space-time corresponding to any given set of multipole moments. 
In addition to an implicit algebraic form for the general solution, 
we also give a power series expression for all finite 
sets of multipole moments.
As conjectured by Geroch we prove in the special case of axisymmetry,
that there is a static space-time for any given set of 
multipole moments subject to a (specified) convergence criterion.
We also use this method to confirm a conjecture
of Hern\'andez-Pastora and Mart\'in concerning the 
monopole-quadropole solution.
\end{abstract}
\section{Introduction}
The relativistic multipole moments of vacuum static asymptotically flat space-times
have been defined by Geroch \cite{geroch}, and this definition has been extended by Hansen 
\cite{hansen} to the
stationary case. There are other definitions \cite{beigAPA},
\cite{thorne}, \cite{quevedo} or approaches \cite{fodor}, \cite{hernandez1998}; for instance, Thorne 
\cite{thorne}
has suggested an alternative definition which is equivalent if the
space-time has nonzero mass \cite{gursel}. 

The recursive definition of Geroch (\ref{orgrec}) produces a family of totally symmetric
and trace-free tensors, which are to be evaluated at a certain point. These values will
then provide the moments of the space-time in question. Even in the case of static 
axisymmetric space-times, where all solutions in a sense are known 
(see  \cite{weyl} and section \ref{exmoments} below), the actual calculations of
the tensors in \eqref{orgrec} are non-trivial. 

In \cite{herb}, it was shown how the moments in the axisymmetric case can be obtained
through a set of recursively defined real valued functions $\{f_n\}_{n=0}^\infty$ on $\mathbb{R}$. The moments
are then given by the values $\{f_n(0)\}_{n=0}^\infty$.  In this way,
one can easily calculate `any' desired number of moments. By exploring the conformal freedom
of the construction, it was also shown how all moments could be captured in one
real valued function $y$, where the moments appeared as the derivatives of $y$ at $0$.

In this paper this feature will be explored further. Namely, we will ask the question: which
static axisymmetric space-time corresponds to a given set of multipole moments?
In the case of a finite number of moments, it is possible to explicitly write down
the metric for the space-time in terms of a power series.
Through this, we can, for instance, prove the monopole-quadropole conjecture
by Hern\'andez-Pastora and Mart\'in, \cite{hernandez}. In the general case, we find an implicit
relation for the moment generating function and  a function related to the Weyl solutions.
Using this relation, we can give the precise condition on the moments in order to have a corresponding
physical space-time with these moments. Thus, this proves a conjecture of Geroch \cite{geroch} in the special
case of axisymmetry.

\section{Explicit moments of the axisymmetric Weyl solutions} \label{exmoments}
We consider a static space-time $M$, with $V$ a 3-surface orthogonal to  
the time-like Killing
vector $\xi^a$. It is required that $V$
is asymptotically flat, i.e., if $h_{ab}$ is the positive definite  
induced metric on $V$, there exists a 3-manifold
$\widetilde V$ and a conformal factor $\Omega$ satisfying
\begin{itemize}
\item[(i)]{$\widetilde V = V \cup \Lambda$, \; where $\Lambda$ is a single point}
\item[(ii)]{$\tilde h_{ab}=\Omega^2 h_{ab}$ is a smooth metric on $\widetilde V$}
\item[(iii)]{At $\Lambda$, $\Omega=0, \tilde D_a \Omega =0, \tilde D_a \tilde  
D_b \Omega = 2 \tilde h_{ab}$,}
\end{itemize}
where $\tilde D_a$ is the derivative operator associated with $\tilde  
h_{ab}$.

On $M$, one defines the scalar potential\footnote{Note that we have changed sign compared
to \cite{geroch}. This gives the monopole $m$ rather than $-m$ for the Schwarzschild solution.}
 $\psi=1-\sqrt{-\xi_a \xi^a}$.  
The multipole moments of $M$ are then defined on $\widetilde V$ as certain  
derivatives of the scalar
potential $\tilde \psi=\psi/\sqrt \Omega$ at $\Lambda$. More  
explicitly, following \cite{geroch}, let $\widetilde R_{ab}$ denote
the Ricci tensor of $\widetilde V$, and let $P=\tilde \psi$. Define the  
sequence $P, P_{a_1}, P_{a_1a_2}, \ldots$
of tensors recursively:
\begin{equation} \label{orgrec} P_{a_1 \ldots a_n}=C[
\tilde D_{a_1}P_{a_2 \ldots a_n}-
\frac{(n-1)(2n-3)}{2}\widetilde R_{a_1 a_2}P_{a_3 \ldots a_n}],
\end{equation}
where $C[\ \cdot \ ]$ stands for taking the totally symmetric and  
trace-free part. The multipole moments
of $M$ are then defined as the tensors $P_{a_1 \ldots a_n}$ at  
$\Lambda$.

If, in addition to the requirement that $M$ is static and asymptotically  
flat, we also impose the condition that
$M$ is axisymmetric, all solutions are in principle known as  
corresponding to solutions of the Laplace equation in flat  3-space  
\cite{weyl}. 
The metric can then be written
\begin{equation} \label{M}
ds^2=-e^{2\alpha}dt^2+e^{2(\beta-\alpha)}(dR^2+dZ^2)+R^2 e^{-2 \alpha}d\phi^2,
\end{equation}
where $\alpha=\frac{1}{2}\ln{|\xi_a\xi^a|}$, which vanishes at infinity, is axisymmetric and flat-harmonic with respect to the cylindrical
coordinates $R,Z$ and $\phi$. Furthermore, $\beta$ is the solution to 
$  \partial_R \beta  =  R[(\partial _R \alpha)^2-(\partial_Z \alpha)^2]$,
$  \partial_Z \beta  =  2 R  (\partial _R \alpha)(\partial _Z \alpha)$,
$  \partial_\phi \beta  =  0$
which vanish at infinity. 
The metric on $V$ is $ds^2=e^{2(\beta-\alpha)}(dR^2+dZ^2)+R^2 e^{-2 \alpha}d\phi^2$, and
to conformally compactify $V$ and introduce suitable coordinates on $\widetilde V$, we
let, \cite{geroch}, \cite{herb}, $\rho=\frac{R}{R^2+Z^2}, \,  z=\frac{Z}{R^2+Z^2}$ and define the spherical coordinates
$r,\theta, \phi$ via $\rho=r \sin\theta,\, z=r\cos\theta, \phi=\phi$. 
Choosing as conformal factor $\Omega=e^{\alpha-\beta}/(R^2+Z^2)$, 
the metric for $\widetilde V$ in a neighbourhood
of $\Lambda$ (which is now the origin point) is 
\begin{equation} \label{rescmet}
d\tilde s^2=
e^{-2\beta} r^2 \sin^2 \theta d \phi^2
+(dr^2+r^2 d\theta^2)
\end{equation}
In terms of the spherical coordinates $r,\theta,\phi$, the function\footnote{With this 
choice of sign, the Schwarzschild solution will also have positive Weyl moments.}
$\widetilde \alpha=-\alpha/r$ is harmonic and $\beta$ satisfies
$(r \partial_r - i \partial_\theta)\beta=
i \sin \theta e^{i\theta}
[(r \partial_r - i \partial_\theta)\alpha]^2$.
%\begin{equation} \label{betaeq}
%\end{equation}
The conformal factor $\Omega$ is not uniquely determined. One can make a further
conformal transformation of $\widetilde V$, using as conformal factor 
$e^{\kappa}$, where
$\kappa$ is any smooth function on $\widetilde V$ with  
$\kappa(\Lambda)=0$. Thus
$\kappa$ reflects the freedom in
choosing $\Omega$. Of particular importance is the
value of $\kappa'(0)$. Namely, under a change
$\Omega \to \Omega e^\kappa$, a non-zero $\kappa'(0)$ changes
the moments defined by (\ref{orgrec}) in a way which
corresponds to a `translation' of the physical space \cite{geroch}. 
The potential $P$ is
\begin{equation} \label{rescpot}
P=\tilde \psi  
=\psi/\sqrt{\Omega}=\frac{e^{(\beta-\kappa)/2}}{r}(e^{-\alpha/2}- 
e^{\alpha/2}).
\end{equation}
For the multiple moments $P_{a_1 \ldots a_n}(\Lambda)$,
the axisymmetry implies that \cite{geroch} at $\Lambda$, $P_{a_1  
\ldots a_n}$ is proportional to
$C[z_{a_1} z_{a_2} \cdots z_{a_n}]$, where $z_a=(dz)_a$, so that
\begin{equation} \label{scalars}
  P_{a_1 \ldots a_n}(\Lambda)=m_n C[z_{a_1} z_{a_2} \cdots z_{a_n}],  
\, \, n\geq 1, \quad m_0=P(\Lambda).
\end{equation}
Thus, in the Weyl case, the moments are given by the sequence
$(m_0, m_1, m_2, ...)$.
As shown in \cite{herb}, all these moments can be collected into one  
single function
$y:\mathbb{R}^+ \cup {0} \to \mathbb{R}$, where the moments appear as  
derivatives
of $y$ at $0$. \\
This is possible due to the form of $\tilde\alpha$ ($P_n$ being the Legendre polynomials)
\begin{equation*}
\tilde\alpha(r,\theta)=\sum_{n=0}^{\infty}a_n r^n P_n(\cos\theta)
\label{aform}
\end{equation*}
together with the form of $C[z_{a_1} z_{a_2} \cdots z_{a_n}]$ at  
$\Lambda$.
We can also write $\beta(r,\theta)$ explicitly as
$\beta=\sum_{k=0}^\infty b_k(\cos\theta) r^{k+2}$, with, c.f. \cite{hernandez},
\begin{equation*}
b_k(z)=\sum_{l=0}^{k}a_l a_{k-l}\tfrac{(l+1)(k-l+1)}{k+2}[P_{l+1}(z)P_{k-l+1}(z)-P_{l}(z)P_{k-l}(z)].
\end{equation*}
As explained in \cite{herb}, to determine the moments
$m_n$, it is sufficient to follow the leading  
order terms
of $P_n$ in the expansion (\ref{aform}). The leading term of
$P_n$ is\footnote{$(a)_n=a(a+1)(a+2)\ldots (a+n-1)=\Gamma(a+n)/\Gamma(a)$}
 $\frac{(2n)!}{2^n n!^2}=\frac{2^n (\frac{1}{2})_n}{n!}$. Therefore, in conjunction with (\ref{aform}),
we also define
\begin{equation} \label{ydef}
Y(r)=\sum_{n=0}^{\infty}a_n r^n \tfrac{(2n)!}{2^nn!^2}.
\end{equation}

We can now state the following theorem, taken from \cite{herb},

\begin{theorem} \label{herbthm}
Suppose that a static axisymmetric asymptotically flat space-time $M$ is given by the flat-harmonic function $\alpha$, 
which after conformal rescaling is given by (\ref{aform}). Let $Y$ be given by (\ref{ydef}), put
$\beta=\left (Y+2r\frac{d Y}{d r}\right)^2$, $h(r)=\int_0^r 2r\beta(r)\,dr$ and define $\kappa$ through
$$\kappa(r)=-\ln \left(-r\int\frac{e^{h(r)}}{r^2}\,dr \right).$$
Put $\rho(r)=r e^{\kappa(r)}$ 
and define $y:\mathbb{R}^+ \cup {0} \to \mathbb{R}$ implicitly by
$y(\rho)=e^{-\kappa(r)/2} Y(r)$.
Then the multipole moments $m_0, m_1, m_2,\ldots$ of $M$ are given by $m_n=\frac{d^n y}{d \rho^n}(0)$.
\end{theorem}

In the definition of $\kappa$, there appears a constant of integration.
This constant affects $\kappa^\prime(0)$. In particular, one can make
$\kappa'(0)=0$.

\subsection{The Schwarzschild solution} \label{schsol}
In this section we illustrate how Theorem \ref{herbthm} can be used to calculate
the moments of the Schwarzschild solution.
As is well known, the Weyl monopole, i.e., $\alpha \equiv 1$ does not correspond
to the Schwarzschild solution. Rather, for Schwarzschild, the corresponding
Weyl potential is
\begin{equation} \label{salpha}
\tilde\alpha(r,\theta)=
\frac{1}{2r}\ln\frac{\sqrt{1+2 r m \cos \theta+r^2 m^2}+\cos \theta + r m}
{\sqrt{1-2 r m \cos \theta+r^2 m^2}+\cos \theta - r m}
\end{equation}
so that 
$
\tilde\alpha(r,0)=\sum_{n=0}^\infty a_n r^n=\frac{1}{2r}\ln\frac{1+mr}{1-mr}=
\sum_{n\geq 0 ,\, \mbox{odd}}^{\infty}\frac{m^n}{n} r^{n-1}
$ and consequently
$
Y(r)=\frac{1}{r}\sum_{n\geq 0, \, \mbox{odd}}^{\infty}
\frac{m^n}{n} r^n 
\frac{(2n-2)!}{2^{n-1}(n-1)!^2}={\frac {m\sqrt {2}}{\sqrt {1+\sqrt {1-4\,{(mr)}^{2}}}}}.
$
One can now follow the steps in Theorem \ref{herbthm} and get 
$\beta(r) =\frac{\sqrt{1-4m^2r^2}-1}{2r^2(4m^2r^2-1)}$,
$h(r) =\ln\left (\frac{1+\sqrt{1-4m^2r^2}}{2\sqrt{1-4m^2r^2}}\right )$,
$\kappa(r) =-\ln\left ( \frac{1}{2}+\frac{\sqrt{1-4m^2r^2}}{2}-r\kappa'(0) \right )$,
$\rho(r) =\frac{2r}{1-2\kappa'(0)+\sqrt{1-4m^2r^2}}$,\\
$r(\rho) =\frac{\rho(\rho\kappa'(0)+1)}{m^2\rho^2+(\rho\kappa'(0)+1)^2}$, and finally
\begin{equation} \label{ysch}
y(\rho) =\frac{m}{\sqrt{\rho\kappa'(0)+1}}.
\end{equation}
With $\kappa'(0)=0$ in (\ref{ysch}), so that $y(\rho) \equiv m$, it is evident that we have the
Schwarzschild solution. This corresponds to an expansion `around the centre of mass' (where
the dipole moment vanishes). A non-zero $\kappa'(0)$ gives a `translated expansion' as explained
in \cite{geroch}.

If we just want to calculate any finite number of moments, it is usually easier
to use some of the other algorithms from \cite{herb}. The strength of Theorem \ref{herbthm}
lies in the fact that all moments are handled simultaneously. This will be used in the next section
where we will address the question of which $\alpha$ 
corresponds to a given set of
multipole moments.

\section{General multipole}
In this section, we will study the following question. 
Given a sequence of multipole moments, together with some
convergence criteria, what is the corresponding space-time?
The moments $\{m_n\}$ are given through the function 
\begin{equation*}
\label{yrho}
y=y(\rho)=\sum_{n=0}^\infty \frac{m_n \rho^n}{n!}
\end{equation*} 
as in Theorem \ref{herbthm}, i.e., through
its derivatives at $0$. We assume that the series converges in some neighbourhood
of $\rho=0$. The space-time is then determined if we know $Y(r)$, since this gives $\alpha$.

We first note from Theorem \ref{herbthm} that $\rho=r e^{\kappa(r)}$. 
This relates $Y$ and $y$ since
 $Y=e^{\kappa/2} y(\rho)= e^{\kappa/2} y(r e^\kappa)$.
 However, this relation is somewhat implicit, and therefore of limited use.

Instead, we write $Y=\sqrt{\frac{\rho}{r}} y(\rho)$ and look for a more direct relation between
$\rho$ and $r$:

\begin{lemma} \label{rhorrel}
If $y$ is given by \eqref{yrho} and $\rho$, $\kappa$ are given implicitly by the relations in Theorem \ref{herbthm}, then 
\begin{equation} \label{rhoekv}
0=-\frac{\rho}{r}+\rho \int_0^\rho \frac{1}{\sigma^2} \int_0^\sigma 2\rho \,(y+2\rho y_\rho)^2\,d\rho
\, d\sigma+1+\kappa'(0)\rho.
\end{equation}
\end{lemma}

\noindent {\bf Proof.} From Theorem \ref{herbthm} we have that 
$r/\rho=e^{-\kappa}=
-r \int \frac{e^{h(r)}}{r^2}\,dr
$ so that 
$e^{h(r)}=r^2\rho_r/\rho^2$
and 
$h(r)=\int_0^r 2 r \beta(r)\, dr
=2\ln r-2\ln \rho+\ln \rho_r$ and consequently
\begin{equation} \label{beta}
2 r \beta = \frac{2}{r}-\frac{2 \rho_r}{\rho}+\frac{\rho_{rr}}{\rho_r},
\end{equation}
where $\beta=(Y+2 r Y_r)^2$.
Putting $G = \sqrt{r} Y(r) = \sqrt{\rho} y(\rho)$, we get
\begin{equation} \label{gr}
G_r^2= \frac{1}{4r}(Y(r)+2r  Y_r)^2=\frac{\beta}{4r} \quad
\mbox{and also} \quad
G_r^2= \frac{1}{4\rho}(y(\rho)+2\rho y_\rho)^2 \rho_r^2,
\end{equation}
where the second expression follows from the chain rule, $G_r=G_\rho \rho_r$.
Eliminating $\beta$ and $G_r$ in \eqref{beta} and \eqref{gr}, we get
$
\frac{2\rho^2}{r^3 \rho_r^2}-\frac{2 \rho}{r^2 \rho_r}+\frac{\rho^2 \rho_{rr}}{r^2 \rho_r^3}=
2 \rho (y+2\rho y_\rho)^2.
$
However, the left hand side is just $-\frac{d}{d\rho}(\frac{\rho^2}{r^2 \rho_r})$, so that
$\frac{\rho^2}{r^2 \rho_r}=-\int_0^\rho 2 \rho (y+2\rho y_\rho)^2\,d\rho+C_1$. From 
$\rho=r e^{\kappa(r)},\, \kappa(0)=0$, the limit $r \to 0$ gives $C_1=1$. Thus,
\begin{equation} \label{innanint}
\frac{1}{r^2 \rho_r}=-\frac{d}{d \rho} (\frac{1}{r})=
-\frac{1}{\rho^2} \int_0^\rho 2 \rho (y+2\rho y_\rho)^2\,d\rho+\frac{1}{\rho^2}.
\end{equation}
A further integration of (\ref{innanint}) followed by a multiplication with $\rho$ yields
$$
0=-\frac{\rho}{r}+\rho \int_0^\rho \frac{1}{\sigma^2} \int_0^\sigma 
2\rho \, (y+2\rho y_\rho)^2\,d\rho
\, d\sigma+1+C_2\rho.
$$
By differentiating with respect to $r$, observing that
\[\rho\frac{d}{dr}\int_0^\rho \frac{1}{\sigma^2} \int_0^\sigma 
2\rho \, (y+2\rho y_\rho)^2\,d\rho
\, d\sigma=\frac{1+r\kappa'-e^\kappa}{r}\]
and taking the limit $r \to 0$, we find that 
$C_2=\kappa'(0)$ which is usually, but not always, put to 0. This gives Lemma \ref{rhorrel}.\\

We will use this relation extensively in the sequel. For instance, we notice that for
a space-time with only a finite number of non-zero moments, $y$ will be a polynomial
in $\rho$. This means that (\ref{rhoekv}) will be an algebraic equation which can be solved
by the techniques presented in section \ref{sae}. For instance, the Schwarzschild solution
is given by the constant function $y(\rho) =m$, so that \eqref{rhoekv}, with $\kappa'(0)=0$ gives
$$\rho^2 m^2-\frac{\rho}{r}+1=0, $$ 
i.e., 
$\rho=\frac{2 r}{1+\sqrt{1-4 m^2 r^2}}$,
$Y=\sqrt{\frac{\rho}{r}}\, y=\frac{\sqrt{2} m}{\sqrt{1+\sqrt{1-4 m^2 r^2}}}$
from which (cf. section \ref{schsol})
$$\tilde\alpha(r,0)=\frac{1}{2r}\ln\frac{1+m r}{1-m r}.$$

\subsection{On a multipole conjecture due to Geroch}
In \cite{geroch}, it was conjectured that
\begin{quote}
Two static solutions of Einstein's equations having identical multipole moments coincide, 
at least in some neighbourhood of $\Lambda$. 
\end{quote}
and that
\begin{quote}
Given any set of multipole moments, subject to the appropriate convergence condition, there exists a static solution of Einstein's equations having precisely those moments. 
\end{quote}
The  first conjecture was proven in \cite{beig}, at least for static space-times with non-zero mass.
We will prove the second conjecture for the special case of axisymmetric space-times
and provide the precise condition on the moments.
\begin{theorem}
Given any set of multipole moments $\{m_n\}_{n=0}^\infty$ defined by (\ref{scalars}), such that 
\begin{equation}
\label{yfromm}
y(\rho)=\sum_{n=0}^\infty{\frac{m_n}{n!}\rho^n}
\end{equation}
converges in some neighbourhood of $\rho=0$, 
there exists an axisymmetric  static solution of Einstein's equations having precisely those moments. Conversely, to every axisymmetric space-time given by $\widetilde{\alpha}(r,0)$ in 
(\ref{aform})  with the corresponding set  of moments $\{m_n\}_{n=0}^\infty$,  the right hand side
of (\ref{yfromm}) converges  in a neighbourhood of $\rho=0$.
\end{theorem}

\noindent {\bf Proof.} 
Suppose $\{m_n\}_{n=0}^\infty$ is such that $y(\rho)=\sum_{n=0}^\infty{\frac{m_n}{n!}\rho^n}$
converges in some neighbourhood of $\rho=0$. From \eqref{rhoekv} we find that
$r(\rho)=\rho/(1+\rho f(\rho))$,
where $f(\rho)=\int_0^\rho \frac{1}{\sigma^2} \int_0^\sigma 2\rho \,(y+2\rho y_\rho)^2\,d\rho
\, d\sigma+\kappa'(0)$ is analytic in a neighbourhood of $\rho=0$. From $r(0)=0$,
$r^\prime(0) \neq 0$, the inverse function theorem for analytic functions \cite{rudin}
gives that $\rho=\rho(r)$ is also an analytic function, with $\rho(0)=0$, in some neighbourhood of $r=0$.
Thus $Y(r)=\sqrt{\rho/r}\,y(\rho)=\sqrt{1+\rho(r) f(\rho(r))}y(\rho(r))$
is analytic in some neighbourhood of $r=0$. The same condition will then hold for
$\widetilde \alpha (r,0)$, so that $\widetilde \alpha(r,\theta)$ is (real) analytic in a 
neighbourhood of $r=0$.

Conversely, suppose $\widetilde \alpha(r,\theta)$ is flat-harmonic in a neighbourhood of $r=0$.
Then $Y$, given by ({\ref{ydef}}), will also be analytical in a neighbourhood of $r=0$ and from
Theorem \ref{herbthm} it is evident that also $\beta$ and $h$ will have this property. Note
that $h(0)=h^\prime(0)=0$, which implies that $\kappa$ is also analytic in a 
neighbourhood of $r=0$ and satisfies $\kappa(0)=0$. From 
$\rho=r e^{\kappa(r)}$, we have $\rho(0)=0$, $\rho^\prime(0) \neq 0$ and that consequently
$r$ is an analytic function of $\rho$ near $\rho=0$ with $r(0)=0$. This implies that
$y(\rho)=e^{-\kappa(r(\rho))/2} Y(r(\rho))$ is analytic in a neighbourhood of $\rho=0$, so that
$y(\rho)=\sum_{n=0}^\infty c_n \rho^n$ converges there. From 
$m_n=\frac{d^n y}{d \rho^n}(0)=c_n\, n!$ the statement follows.

\subsection{Finite number of moments}
In the special case of a finite number of (non-zero) moments, the function $y$ will be a polynomial
in $\rho$. This implies that Lemma \ref{rhorrel} will give an algebraic relation between
$\rho$ and $r$. Regarding $\rho$ as the unknown, we will have a polynomial equation
where the coefficients depends on the desired moments and $r$. 
Namely, suppose $y(\rho)=\sum_{k=0}^{k=n} \frac{m_k}{k!}\rho^k$ so that all moments
$m_{n+1}, m_{n+2}, \ldots$ are zero. From (\ref{rhoekv}) we find that $\rho$ satisfies the equation
\begin{equation} \label{ekvform}
a_{2n} \rho^{2n}+a_{2n-1} \rho^{2n-1}+ \cdots +a_2 \rho^2 + 
(\kappa^\prime(0)-\frac{1}{r}) \rho+1=0
\end{equation}
so that $a_0=1$, $a_1=\kappa^\prime(0)-\frac{1}{r}$. All other coefficients
$a_2, a_3, \ldots a_{2n}$ are polynomial expressions in the moments $m_0, m_1, \ldots m_n$.
As we will see in section 
\ref{sae}, this equation can be solved in terms of a power series, and this will enable us to write down
$Y$ and $\alpha$ explicitly in terms of a power series.

\section{Solving algebraic equations} \label{sae}
If we have a finite number of moments, equation \eqref{rhoekv} reduces  
to the algebraic equation (\ref{ekvform}).
In this section we therefore study solutions to a general algebraic  
equation
\begin{equation}
\label{algeq}
a_n X^n+a_{n-1} X^{n-1}+\dots+a_1 X+a_0=0.
\end{equation}
This equation can be solved in terms of $\mathcal{A}$-hypergeometric series
in the sense of Gel'fand, Kapranov and Zelevinsky, \cite{GKZ}. Sturmfels \cite{sturmfels}
constructs the solutions explicitly in the form of power series. 
Loosely speaking, each of the $n$ solutions to (\ref{algeq}) can be written down as 
power series in several ways, depending on how one combines the coefficients $a_k$.
For our purpose, we seek the solution $\rho$ to (\ref{ekvform}) where
$\rho/r \to 1$ as $r \to 0$. Furthermore, since we are looking for solutions which converge
near $r=0$, we note that in (\ref{ekvform}), $a_1$ is `large'.

To be consistent with \cite{sturmfels} we introduce the following notations for $u \in \mathbb{Q}$,
$v \in \mathbb{Z}$

\begin{equation*}
\mathcal{A}=
\begin{pmatrix}
0 & 1 & \dots & n-1 & n\\
1 & 1 & \dots & 1   & 1
\end{pmatrix}
\end{equation*}
\begin{equation*}
\gamma(u,v) =
\begin{cases}
1 & v=0\\
u(u-1)\dots (u+v+1) & v <0\\
0 & 0 > u \geq -v, u\in \mathbb{Z}\\
\tfrac{1}{(u+1)(u+2)\dots(u+v)} & \text{otherwise}
\end{cases}
\end{equation*}
Note that if $u$ is not a negative integer then  
$\gamma(u,v)=\Gamma(u+1)/\Gamma(u+v+1)$.
We also define the formal power series
\begin{equation*}
[a_0^{u_0}a_1^{u_1}\dots a_n^{u_n}] = \sum_{(v_0,\dots,v_n)\in  
\mathcal{L}}\prod_{i=0}^n(\gamma(u_i,v_i)a_i^{u_i+v_i}),
\end{equation*}
where $\mathcal{L}$ is the integer kernel of $\mathcal{A}$,
i.e., integer solutions to the equation $\mathcal{A} \overline v=0$.
Sturmfels \cite{sturmfels} then states that $-[a_0a_1^{-1}]$ is a solution to  
\eqref{algeq}:
\begin{equation*}
X=-[a_0a_1^{-1}]=-\sum_{v_2,\dots,v_n\geq 0}{a_0^{1+v_0}a_1^{-1+v_1}
\frac{\gamma( 
-1,v_1)}{\Gamma(2+v_0)}\prod_{j=2}^n{\frac{a_j^{v_j}}{v_j!}}},
\end{equation*}
where
$v_1=-\sum_{i=2}^n{iv_i}$, 
$v_0=\sum_{i=2}^n{(i-1)v_i}$.
We also note that for negative integers $v_1$, 
$\gamma(-1,v_1)=\prod_{i=0}^{-v_1-1}(-1-i)=(-1)^{v_1}\Gamma(1-v_1)$,
which leads to the following expression for the solution
\begin{equation}
\label{solalgeq}
X = \left ( \frac{-a_0}{a_1} \right ) \sum_{v_2\dots v_n \geq 0}
{\frac{a_0^{\sum_{j=2}^n{(j-1)v_j}}  
(\sum_{j=2}^n{jv_j})!}{(-a_1)^{\sum_{j=2}^n{jv_j}}  
(1+\sum_{j=2}^n{(j-1)v_j)!}}\prod_{j=2}^n{\frac{a_j^{v_j}}{v_j!}}}
\end{equation}
Despite its appearance, \eqref{solalgeq} is useful, since an application to \eqref{ekvform},
with $\kappa^\prime(0)=0$, yields $\rho$ as a power series in terms of $r$. The issue of
convergence in \eqref{solalgeq} will be addressed in section \eqref{powers}.

\section{Powers of the root}  \label{powers}
With $\rho$ known in terms of $r$, we can almost write down the function
$Y=\sqrt{\rho/r} y(\rho)$ where $y$ is a polynomial. However, we need to
take fractional powers of $\rho$. Powers of multiple series  
are cumbersome, but in this case they can be expressed explicitly by the following extension
of the work in \cite{sturmfels}. We begin with a lemma describing 
fractional powers of \eqref{solalgeq} in terms of formal power series. 
We then extend this to real (positive) powers and prove convergence.

\begin{lemma} \label{Xpotens}
Let $X$ be the root of \eqref{algeq} given by \eqref{solalgeq} and assume that $\gamma\in\mathbb{Q}^+\backslash\mathbb{Z}$. Then 
\begin{equation}
\label{solpower}
X^\gamma = \left ( \frac{-a_0}{a_1} \right )^\gamma \sum_{v_2\dots v_n  
\geq 0}
{\frac{\gamma a_0^{\sum_{j=2}^n{(j-1)v_j}}  
\Gamma(\gamma+\sum_{j=2}^n{jv_j})}{(-a_1)^{\sum_{j=2}^n{jv_j}}  
\Gamma(1+\gamma+\sum_{j=2}^n{(j 
-1)v_j)}}\prod_{j=2}^n{\frac{a_j^{v_j}}{v_j!}}}.
\end{equation}
\end{lemma}

\noindent{\bf Proof.}
We first see \cite{sturmfels} that a solution $X$ to \eqref{algeq} satisfies
\begin{equation} \label{Xsystem}
\begin{split}
&\sum_{i=0}^n{ia_i\dfrac{\partial X}{\partial a_i}} = -X, \quad 
\sum_{i=0}^n{a_i\dfrac{\partial X}{\partial a_i}} = 0,\\
&\dfrac{\partial^2X}{\partial a_i \partial a_j} =  
\dfrac{\partial^2X}{\partial a_k \partial a_l}
\quad \mbox{whenever} \quad i+j=k+l.
\end{split}
\end{equation}
This implies (c.f. the proof for (\ref{Xsystem}) in \cite{sturmfels}) that $Y=X^\gamma$ satisfies the following system of  
equations
\begin{equation} \label{pdesystem}
\begin{split}
&\sum_{i=0}^n{ia_i\dfrac{\partial Y}{\partial a_i}}  = -\gamma Y, \quad 
\sum_{i=0}^n{a_i\dfrac{\partial Y}{\partial a_i}} = 0,\\
&\dfrac{\partial^2 Y}{\partial a_i \partial a_j} =  
\dfrac{\partial^2 Y}{\partial a_k \partial a_l}
\quad \mbox{whenever} \quad i+j=k+l.
\end{split}
\end{equation}
The assumption $\gamma \in \mathbb{Q^+}\backslash\mathbb{Z}$ together with
Lemma 3.1 in Sturmfels \cite{sturmfels} implies that $[a_0^\gamma a_1^{-\gamma}]$  
satisfies the system \eqref{pdesystem}, and homogeneity implies that 
$\tilde{X}=(-1)^\gamma[a_0^\gamma a_1^{-\gamma}]$ also satisfies the  
same system. Written out explicitly this means that
\begin{equation*}
\tilde{X}
=(-1)^\gamma \sum_{v_2,\dots,v_n\geq  
0}{a_0^{\gamma+v_0}a_1^{-\gamma+v_1}
\frac{\Gamma(\gamma+1)\Gamma(1-\gamma)}{\Gamma(\gamma+v_0+1)\Gamma(1- 
\gamma+v_1)}\prod_{j=2}^n{\frac{a_j^{v_j}}{v_j!}}}.
\end{equation*}
solves \eqref{pdesystem}.
In the general case, the occurrence of the fractional power $\gamma$ means that some care has to be taken
when choosing branches. In our case however, all coefficient $a_i$ are real, and in particular 
$(-a_0/a_1)$ is positive if $r$ is small enough. Thus, no resulting ambiguity will occur from
$(-a_0/a_1)^\gamma $.
Using the relations
$\Gamma(1-\gamma+v_1)=\frac{\pi}{\sin(\gamma\pi-v_1\pi)\Gamma(\gamma- 
v_1)}$ and
$\Gamma(1-\gamma)=\frac{\pi}{\sin(\gamma\pi)\Gamma(\gamma)}$,
we write 
\begin{equation*}
\tilde{X}=(-1)^\gamma \sum_{v_2,\dots,v_n\geq  
0}{a_0^{\gamma+v_0}a_1^{-\gamma+v_1}
\frac{\gamma\sin(\gamma\pi-v_1\pi)\Gamma(\gamma- 
v_1)}{\sin(\gamma\pi)\Gamma(1+\gamma+v_0)}\prod_{j=2}^n{\frac{a_j^{v_j}} 
{v_j!}}}.
\end{equation*}
Plugging in $v_1=-\sum_{i=2}^n{iv_i}$ and $v_0=\sum_{i=2}^n{(i-1)v_i}$,
we see that $\tilde X$ has the form \eqref{solpower}. Thus Lemma \ref{Xpotens} is proven
if we show that $\tilde X=X^\gamma$.

This will follow from Lemma \ref{Ylemma} which says that the solution to \eqref{pdesystem}
of the form \eqref{Yform}
is unique provided certain initial values are given. Both $\tilde X$ and $X^\gamma$
have the form in Lemma \ref{Ylemma} below with $b_0=1$ and $b_1=0$. 
Therefore, $\tilde{X}=X^{\gamma}$ for all $\gamma \in  
\mathbb{Q^+}\backslash \mathbb{Z}$, and the proof is complete.

\begin{lemma} \label{Ylemma}
Assume that a solution to the system \eqref{pdesystem} has the form
\begin{equation}
\label{Yform}
Y=(-\tfrac{a_0}{a_1})^\gamma
\sum_{j=0}^\infty{b_j(a_0,a_2,a_3,\dots,a_n)(-a_1)^{-j}}
\end{equation}
where $b_j$ are analytic and that $\gamma$ is not a negative integer.
Then $Y$ is determined by the values of $b_0$ and $b_1$.
\end{lemma}

\noindent{\bf Proof.}
By linearity we can assume $b_0=b_1=0$ and show that $Y=0$.
Now, let $c_j(a_0,a_2,\dots,a_n)=a_0^\gamma b_j(a_0,a_2,\dots,a_n)$.
Given $c_0=c_1=0$, we will prove by induction that $c_j=0$ for all $j\geq 0$.
Assume that $c_{j-2}=c_{j-1}=0$.
$\frac{\partial^2 Y}{\partial a_1^2}=\frac{\partial^2 Y}{\partial  
a_0\partial a_2}$
gives
$\sum_{k=0}^\infty{(k+\gamma)(k+\gamma+1)(-a_1)^{-k-\gamma-2}c_k}=
\sum_{k=0}^\infty{\frac{\partial^2 c_k}{\partial a_0 \partial a_2}}$.
Identification of the coefficients implies
$\frac{\partial^2 c_j}{\partial a_0 \partial  
a_2}=(j-2+\gamma)(j-1+\gamma)c_{j-2}=0$.
The analyticity of $b_j$ lets us define $d_{j,k}(a_2,a_3,\dots,a_n)$ such that  
$b_j=\sum_{k=0}^\infty{d_{j,k}a_0^k}$. We can then write
$0=\frac{\partial^2 c_j}{\partial a_0 \partial a_2}=
\sum_{k=0}^\infty{(k+\gamma)a_0^{k+\gamma-1}\frac{\partial  
d_{j,k}}{\partial a_2}}$.
Identification of the coefficients implies
$\frac{\partial d_{j,k}}{\partial a_2}=0$.
Hence, $d_{j,k}(a_2,a_3,\dots,a_n)=d_{j,k}(a_3,\dots,a_n)$ for all $k$. From \eqref{pdesystem} we also have
%\begin{equation*}
$
\frac{\partial^2 Y}{\partial a_0\partial a_m}=\frac{\partial^2  
Y}{\partial a_1\partial a_{m-1}},\quad 2<m\leq n,
$
%\end{equation*}
which gives
\begin{equation*}
\frac{\partial^2 c_j}{\partial a_0\partial  
a_m}=(j-1+\gamma)\frac{\partial c_{j-1}}{\partial a_{m-1}}=0,\quad  
2<m\leq n.
\end{equation*}
Using this in the same way as above gives that $d_{j,k}$ does 
not depend on $a_m$, $m=3, 4, \ldots n$. 
Therefore each $d_{j,k}$ is constant. 
Thus,
$$
Y=(-\tfrac{a_0}{a_1})^\gamma
[b_j(a_0) (-a_1)^{-j}+ 
\sum_{k=j+1}^\infty{b_k(a_0,a_2,a_3,\dots,a_n)(-a_1)^{-k}}].
$$
However, 
using $-\gamma Y=\sum_{i=0}^n{ia_i\frac{\partial Y}{\partial  
a_i}}$ from
\eqref{pdesystem} and looking at  the coefficient of $(-a_1)^{-j-\gamma}$,
we see that 
$-\gamma a_0^\gamma b_j(a_0)=-(j+\gamma)a_0^\gamma b_j(a_0)$,
so that $b_j$ and therefore $c_j$ are zero.
Induction gives $c_j=0$ for  
all $j\in \mathbb{N}$. Hence the lemma is proven.

\begin{theorem}
Let $X$ be the root of \eqref{algeq} given by \eqref{solalgeq} and assume that $\gamma\in\mathbb{R}^+$.
Then \eqref{solpower} holds and the series converges if
\begin{equation*}
\left | \frac{a_0}{a_1} \right |^2 < \min\left (
  \frac{1}{ne\sum_{j=2}^n{|\frac{a_j}{a_0}|}},1\right ).
\end{equation*}
\end{theorem}

\noindent{\bf Proof.}
Let $r=-a_0/a_1$, $M=\sum_{j=2}^n{|\frac{a_j}{a_0}|}$ , $N_v=\sum_{j=2}^n{v_j}$, $K_v=\sum_{j=2}^n{jv_j}$, and put
\begin{equation*}
Z(\gamma)=\frac{\tilde X}{\gamma r^\gamma}
=\sum_{v_2,\dots,v_n\geq 0}{\frac{\Gamma(\gamma+K_v)r^{K_v}}{\Gamma(1+\gamma+K_v-N_v)}
\prod_{j=2}^n{\frac{a_j^{v_j}}{a_0^{v_j}v_j!}}},
\end{equation*}
where $\tilde X$ is the right hand side of \eqref{solpower}.
Let $0\leq i \leq \gamma \leq i+1$ and $|r|<1$ and observe that
$2N_v\leq K_v\leq nN_v$. We then have
\begin{align*}
|Z(\gamma)|&\leq \sum_{v_2,\dots,v_n\geq 0}{\frac{(i+K_v)!|r|^{K_v}}{(i+K_v-N_v)!}
\prod_{j=2}^n{\frac{|\frac{a_j}{a_0}|^{v_j}}{v_j!}}}\\
 &\leq \sum_{v_2,\dots,v_n\geq 0}{(i+nN_v)^{N_v}|r|^{2N_v}
\prod_{j=2}^n{\frac{|\frac{a_j}{a_0}|^{v_j}}{v_j!}}}\\
 &=\sum_{N=0}^\infty{\frac{(i+nN)^N|r|^{2N}}{N!}
 \sum_{v_2+\dots+v_n=N}{
N!\prod_{j=2}^n{\frac{|\frac{a_j}{a_0}|^{v_j}}{v_j!}}}}\\
 &=\sum_{N=0}^\infty{\frac{(i+nN)^N|r|^{2N}M^N}{N!}}
=\sum_{N=0}^\infty{b_N}.
\end{align*}
%\begin{align}
%\frac{b_{N+1}}{b_N}&=\frac{(N+1)^N(\frac{i}{N+1}+n)^{N+1}}{N^N(\frac{i}{N}+n)^N}M|r|^2\\
%&\leq (1+\frac{1}{N})^N(\frac{i}{N}+n)M|r|^2 \rightarrow enM|r|^2
%\end{align}
Furthermore, $b_{N+1}/b_N \to e n M |r|^2$ as $N \to \infty$, 
so $Z(\gamma)$ converges uniformly on $\gamma\in [i,i+1] $ if $|r| <
\frac{1}{\sqrt{nMe}}$ and $|r| < 1$. From the uniform convergence it 
follows that $Z(\gamma)\in C(\mathbb{R}^+)$, since 
$\frac{\Gamma(\gamma+K_v)}{\Gamma(1+\gamma+K_v-N_v)}\in
C(\mathbb{R}^+)$ and $i\in \mathbb{N}$ is arbitrary. 
$X^{\gamma}$ is continuous with respect to $\gamma$, 
and since $X^{\gamma}$ equals $\tilde X= \gamma r^\gamma Z(\gamma)$ 
on a dense subset of $\mathbb{R}^+$, the theorem follows.

{\bf Remark.} When $i=0$, we should have $0 <  \gamma \leq 1$, and we also note that
the estimates in the proof fail for one term. This does not affect the conclusion.

%Thus the right hand side in the theorem is continuous. The left hand side is also continuous. 
%They are equal on a dense subset of $\mathbb{R}^+$ and therefore equal for all $\gamma\in %\mathbb{R}^+$.
%Furthermore the series converges if $|r| < \frac{1}{\sqrt{nMe}}$.

\section{Examples}
In this section we give examples of how to find expressions for $\tilde \alpha$ 
given only the multipole moments. We start with the pure $2^n$-poles which include
the Schwarzschild solution and the gravitational dipole of \cite{herb}. We then give
$\tilde \alpha$ for the monopole -$2^n$-poles, which in particular include the
monopole-quadropole solution of Hern\'andez-Pastora and Mart\'in, \cite{hernandez} 

\subsection{Pure $2^n$-pole}
For the pure $2^n$-pole we have $y(\rho)=q\frac{\rho^n}{n!}$,
so that equation \eqref{rhoekv} reduces to
\begin{equation*}
\frac{(1+2n)q^2}{n!(n+1)!}\rho^{2n+2} -\frac{\rho}{r} +1 =0.
\end{equation*}

The solution $\rho(r)$ with the appropriate asymptotics is given by \eqref{solalgeq},
while
$Y(r)=\frac{q\rho^{n+1/2}}{\sqrt{r}n!}$
 is given by \eqref{solpower}.
The resulting power series expression is
\begin{equation*}
Y(r)=\frac{r^nq}{n!}\sum_{i=0}^\infty{\frac{(2n+1)\Gamma(2(n+1)i+n+1/2)r^{2(n+1)i}c^i}{2\Gamma((2n+1)i+n+3/2)i!}},
\end{equation*}
where 
$c=\frac{(1+2n)q^2}{n!(n+1)!}$.
In terms of $\tilde \alpha$ we get
\begin{equation}
\label{purealpha}
\tilde\alpha(r,\theta)=\frac{(2n+1)q}{n!}\sum_{i=0}^\infty{\frac{(2(n+1)i+n)!\sqrt{\pi}c^ir^{2(n+1)i+n}P_{2(n+1)i+n}(\cos\theta)}
{\Gamma((2n+1)i+n+3/2)i!2^{2(n+1)i+n+1}}}.
\end{equation}

%\begin{equation}
%\tilde\alpha(r,0)=\frac{qr^n}{(2n-1)!!}\sum_{i=0}^\infty{\frac{\prod_{j=0}^{2n+1}\left (\frac{n+1+j}{2n+2} \right)_i }
%{i!\prod_{j=0}^{2n}\left( \frac{2n+3+2j}{4n+2}\right )_i} 
%\left (\frac{(n+1)^{2n+1}q^2r^{2n+2}}{(2n+1)^{2n}n!^2} \right )^i}
%\end{equation}

This simple form makes it possible to express $\tilde\alpha(r,0)$ in terms of a hypergeometric function:
\begin{equation}
\label{purealphanoll}
\tilde\alpha(r,0)=\tfrac{qr^n}{(2n-1)!!}
{}_{2n+2}F_{2n+1}([\tfrac{n+1}{2n+2},\tfrac{n+2}{2n+2},\ldots,\tfrac{3n+2}{2n+2}]
,[\tfrac{2n+3}{4n+2},\tfrac{2n+3+2}{4n+2},\ldots,\tfrac{6n+3}{4n+2}],z),
\end{equation}
where $z=\tfrac{(n+1)^{2n+1}q^2}{(2n+1)^{2n}n!^2}r^{2n+2}$.
In particular, \eqref{purealphanoll} with $n=0$ with $q=m$ gives
\begin{equation*}
\tilde\alpha(r,0)=m\,
{}_{2}F_{1}([\tfrac{1}{2},\tfrac{2}{2}]
,[\tfrac{3}{2}],m^2 r^2)=\tfrac{1}{2}\log \tfrac{1+mr}{1-mr},
\end{equation*}
i.e., the Schwarzschild solution.  $n=1$ gives
\begin{equation*}
\tilde\alpha(r,0)=q r
\, {}_{4}F_{3}([\tfrac{2}{4},\tfrac{3}{4},\tfrac{4}{4},\tfrac{5}{4}]
,[\tfrac{5}{6},\tfrac{7}{6},\tfrac{9}{6}],\tfrac{8}{9}m^2 r^4),
\end{equation*}
i.e., the gravitational dipole of \cite{herb}.

\subsection{Monopole - $2^n$-pole}
Space-times with given monopole-moment $m$ and $2^n$-pole-moment $q$
have direct physical interpretations. They describe pure relativistic $2^n$-pole
corrections to the Schwarzschild solution. Of particular interest is the 
monopole-quadropole solution, since it in a sense\footnote{If the mass is non-zero,
the dipole moment can always be put to zero.}
 represents the lowest order correction. This case is considered in section (\ref{mqc}).
In this section we derive the the metric for the general monopole - $2^n$-pole.

A space-time with monopole $m$ and $2^n$-pole $q$ is given by
$
y(\rho)=m+\tfrac{q}{n!}\rho^n.
$
Equation \eqref{rhoekv} then reduces to
\begin{equation*}
c_1\hat q^2\hat\rho^{2n+2}+c_2\hat q\hat\rho^{n+2}+\hat\rho^2-\frac{\hat\rho}{\hat r}+1=0,
\end{equation*}
where $c_1 = \frac{2n+1}{(n+1)!n!}$, $c_2 = \frac{8n+4}{(n+2)!}$,
$\hat\rho = m\rho$, $\hat q = m^{-(n+1)}q$, and $\hat r = mr$.
The solution with the appropriate asymptotics is then given by \eqref{solalgeq}, 
while the $\gamma$-power is given by \eqref{solpower}. The solutions simplifies to
\begin{align*}
\hat\rho &= \hat r \sum_{i,j,k \geq  
0}{\frac{\Gamma(F_{nijk}+1)\hat q^{j+2k}c_2^jc_1^k}{i!j!k! 
\Gamma(G_{nijk}+1)}\hat r^{F_{nijk}}},\\
\hat \rho^\gamma &= \hat r^\gamma \gamma \sum_{i,j,k \geq  
0}{\frac{\Gamma(F_{nijk}+\gamma)\hat q^{j+2k}c_2^jc_1^k}{i!j!k! 
\Gamma(G_{nijk}+\gamma)}\hat r^{F_{nijk}}},
\end{align*}
where we have used the index functions
$F_{nijk}=2i+(n+2)j+(2n+2)k$,
$G_{nijk}=i+(n+1)j+(2n+1)k+1$.
The function $Y$ is then given by
\begin{align*}
Y=&\sqrt{\frac{\rho}r}y(\rho)=\frac{m\hat\rho^{1/2}}{\hat r^{1/ 
2}}+\frac{m\hat q\hat \rho^{n+1/2}}{n!\hat r^{1/2}}\\
=&
\sum_{i,j,k \geq  
0}{
\frac{m\hat q^{j+2k}c_2^jc_1^k\hat r^{F_{nijk}}}{i!j!k! 
}
\left (
\frac{\Gamma(F_{nijk}+\tfrac{1}{2})}{2\Gamma(G_{nijk}+\tfrac{1}{2})} 
+ \frac{(2n+1)\hat q\hat r^n\Gamma(F_{nijk}+n+\tfrac{1}{2})}{2\Gamma(G_{nijk}+n+\tfrac{1}{2})n!}
\right )}.
\end{align*}
After some simplifications we get the following expression for $\tilde\alpha$:
\begin{equation}
\label{alphasolexp}
\begin{split}
\tilde\alpha(r,\theta)= \sum_{i,j,k \geq  
0}{\frac{m\hat q^{j+2k}c_2^jc_1^k\hat r^{F_{nijk}}}{i!j!k! 
2^{i+j+k}}
\left ( \frac{
F_{nijk}!
P_{F_{nijk}}(\cos\theta)}{
(2G_{nijk}-1)!! }\right . }\\
+ \left . \frac{(2n+1)
\hat q\hat r^n
(F_{nijk}+n)! 
P_{F_{nijk}+n}(\cos\theta)}
{n!(2G_{nijk}+2n-1)!!} \right ),
\end{split}
\end{equation}
so that the corresponding space-time is now explicitly given.

\subsection{The monopole-quadropole conjecture of Hern\'andez-Pastora and  
Mart\'in} \label{mqc}
The monopole-quadrupole
(static axisymmetric) solution, i.e., the monopole-$2^2$-pole has been studied
by Hern\'andez-Pastora and Mart\'in.
In \cite{hernandez} they
conjectured an expression for this solution with mass parameter $m$ and quadrupole moment $q$.
With a minor change of conventions (sign and $3q \rightarrow q$), the conjectured solution is
\begin{equation}
\label{hpm}
\tilde\alpha(r,0)= \sum_{i,j,k}
{\frac{m^{2i+1}r^{2i} \hat q^j i!(2i-1)!!3^{-k}5^k(4j-3k+1)}{(2i+2j+1)!!(i-j-k)!(j-k)!(2k-j+1)!}}.
\end{equation}
Here the summation is taken over all integers $i,j,k$ subject to the conditions
$i \geq 0$, $i-j-k \geq 0$, $j-k \geq 0$, $2k-j+1 \geq 0$. Adopting the convention that $1/i!=0$ if $i$ is a negative integer, the summation is taken over all integers $i,j,k$ subject to $i \geq 0$.
Also, $(-1)!!=1$.
We prove this conjecture by rewriting \eqref{alphasolexp}. From \eqref{alphasolexp} we get
\begin{equation*}
\tilde\alpha(r,0)= \sum_{i,j,k}{\frac{m\hat q^{j+2k}c_2^jc_1^k\hat r^{2i+4j+6k}}{i!j!k! 
2^{i+j+k}}
\left ( \tfrac{(2i+4j+6k)!}{(2i+6j+10k+1)!!}
+ \tfrac{5 \hat q\hat r^2 (2i+4j+6k+2)!}
{2(2i+6j+10k+5)!!} \right )},
\end{equation*}
where $c_2=5/6$, $c_1=5/12$ and where by the convention the summation is effectively taken over all 
nonnegative integers. In the first term, let $i\rightarrow i-2j-3k$ and in the second
$i\rightarrow i-2j-3k-1$.
After this change we let $j\rightarrow j-2k$ in the first term and $j\rightarrow j-2k-1$ in the second, 
and obtain
\begin{equation*}
\tilde\alpha(r,0)= \sum_{i,j,k}
{\frac{m\hat q^jc_2^{j-2k}c_1^k\hat r^{2i}(2i)!}{(2i+2j+1)!!k! 
2^{i-j}}
\left ( \tfrac{1}{(i-2j+k)!(j-2k)!}
+ \tfrac{5}{2c_2(i-2j+k+1)!(j-2k-1)!} \right )}.
\end{equation*}
Next, we let $k\rightarrow k-1$ in the second term and obtain
\begin{equation*}
\tilde\alpha(r,0)= \sum_{i,j,k}
{\frac{m\hat q^j5^{j-k}3^{k-j}\hat r^{2i}(2i)!}{(2i+2j+1)!!k! 
2^{i}(i-2j+k)!}
\left ( \tfrac{1}{(j-2k)!}
+ \tfrac{5k}{(j-2k+1)!} \right )}.
\end{equation*}
Using $(2i)!/2^i=(2i-1)!!i!$ and rewriting the first term we get
\begin{equation*}
\tilde\alpha(r,0)= \sum_{i,j,k}
{\frac{m\hat r^{2i} \hat q^j 5^{j-k}3^{k-j} i!(2i-1)!!(j+3k+1)}{(2i+2j+1)!!k!(i-2j+k)!(j-2k+1)!}}.
\end{equation*}
After the change $k\rightarrow j-k$ we finally get \eqref{hpm}.

\section{Conclusions and discussion}
We have investigated the Geroch moments which are defined through the recursion
(\ref{orgrec}). In \cite{herb}, it was shown how this recursion in the
axisymmetric case can be replaced by a much simpler recursion
involving only scalar valued functions, and how all moments
can be captured in one single function. 

In this work, we have showed how these results can be used to construct static axisymmetric space-times
with prescribed moments. In the case of finite number of non-zero
moments, we have shown how to obtain the metric of the corresponding
space-time explicitly in terms of power series. In addition, using
\eqref{alphasolexp} we
were able to confirm the monopole-quadropole conjecture by Hern\'andez-Pastora and  
Mart\'in \cite{hernandez}.  In the general axisymmetric case, where we have the implicit relation 
(\ref{rhoekv}), we were able to confirm a special case of a conjecture due to Geroch \cite{geroch}.

For further studies of the solutions presented here, integral representations or other analytical 
expressions would be very useful. In particular, this could give
further insight for the pure $2^n$-poles \eqref{purealpha}.

It is natural to try and extend the framework presented here to more general
cases, like static non-axisymmetric or stationary space-times.  For instance, it would
be instructive to calculate the moments of the Kerr solution which are stated in \cite{hansen}.

It is also interesting to note that one can define `addition' of space-times, by simply
adding their Geroch moments. For instance, we can `add'  two Schwarzschild solutions 
using (\ref{ysch}) with different displacements. Displacing the solutions with $\kappa'(0)=\pm 1$ and choosing $m=1$, addition of the Geroch-moments 
gives
$y(\rho)=\frac{1}{\sqrt{1+\rho}}+\frac{1}{\sqrt{1-\rho}}$
and by \eqref{rhoekv} we get
\begin{equation} \label{concleq}
-4\sqrt{1-\rho^2}+3+\frac{2}{1-\rho^2}-\frac{\rho}{r}=0.
\end{equation}
This equation is algebraic in terms of $z=(1-\rho^2)^{-1/2}$, and can therefore
be studied with the tools provided here. Thus, it may be possible to investigate the
physical relevance of such an  addition.

\end{document}